
\input harvmac
\newcount\figno
\figno=0
\def\fig#1#2#3{
\par\begingroup\parindent=0pt\leftskip=1cm\rightskip=1cm\parindent=0pt
\baselineskip=11pt
\global\advance\figno by 1
\midinsert
\epsfxsize=#3
\centerline{\epsfbox{#2}}
\vskip 12pt
{\bf Fig. \the\figno:} #1\par
\endinsert\endgroup\par
}
\def\figlabel#1{\xdef#1{\the\figno}}
\def\encadremath#1{\vbox{\hrule\hbox{\vrule\kern8pt\vbox{\kern8pt
\hbox{$\displaystyle #1$}\kern8pt}
\kern8pt\vrule}\hrule}}

\overfullrule=0pt

%
\def\tilde{\widetilde}
\def\bar{\overline}

\def\S{{\bf S}}
\def\R{{\bf R}}

\font\zfont = cmss10 

\def\bigone{\hbox{1\kern -.23em {\rm l}}} \def\ZZ{\hbox{\zfont Z\kern-.4emZ}}

\Title{ \vbox{\baselineskip12pt
\hbox{hep-th/9510169}
\hbox{IASSNS-HEP-95-81}
\hbox{NSF-ITP-95-135}}}
{\vbox{\centerline{EVIDENCE FOR HETEROTIC - TYPE I}
\bigskip
\centerline{STRING DUALITY}}}
\centerline{Joseph Polchinski\footnote{$^\dagger$}
{joep@itp.ucsb.edu}}
\smallskip\centerline{\it Institute for Theoretical Physics}
\centerline{\it University of California, Santa Barbara, CA 93106}
\smallskip\centerline{and}
\smallskip\centerline{Edward Witten\footnote{$^*$}
{witten@sns.ias.edu}}
\smallskip\centerline{\it School of Natural Sciences, Institute for
Advanced Study}
\centerline{\it Olden Lane, Princeton, N.J. 08540}
\bigskip

\medskip

\noindent
A study is made of the
implications of heterotic string $T$-duality and extended gauge symmetry for
the
conjectured equivalence of heterotic and Type I superstrings. While at first
sight heterotic string world-sheet dynamics appears to conflict with Type I
perturbation theory, a closer look shows that Type I perturbation theory
``miraculously'' breaks down, in some cases via novel mechanisms, whenever the
heterotic string has massless particles not present in Type I perturbation
theory. This strongly suggests that the two theories actually are equivalent.
As further evidence in the same direction, we show that the Dirichlet one-brane
of type I string theory has the same world-sheet structure as the heterotic
string.

\Date{October, 1995}

\bigskip
\noindent{\it Introduction}

Of the various plausible equivalences between superstring theories, one
candidate
for which there has been comparatively little evidence is the relation of
the two
ten-dimensional theories with gauge group $SO(32)$ -- the heterotic string and
the Type I theory. Evidence that they are equivalent consists in part of two
facts \ref\witten{E. Witten, ``String Theory Dynamics In Various Dimensions,''
Nucl. Phys. {\bf B433} (1995) 85. } that enable one to avoid immediate
contradictions: the identification between the low energy limits of the two
theories exchanges weak and strong coupling, putting a disproof of the
conjecture
out of reach of perturbation theory; the range of dimensions in which this
equivalence would determine the strong coupling dynamics of the heterotic
string
(eight through ten) does not overlap with the range (seven and below) in which
equivalence with Type II controls that dynamics, again evading a potential
contradiction.
\nref\dab{Atish Dabholkar, ``Ten Dimensional Heterotic String As A Soliton,''
Phys. Lett. {\bf B357} (1995) 307.} \nref\hull{C. M. Hull, ``String-String
Duality In Ten Dimensions,'' Phys. Lett. {\bf B357} (1995) 545.} Further
suggestive discussions have involved attempts to interpret the heterotic string
as a classical solution of the Type I theory \refs{\dab,\hull}.
In this paper, the issue will be further explored in two directions.

First, following a strategy that has proven fruitful in other cases, we will
consider the relation between the self-dualities and the gauge symmetries of
the
two theories when toroidally compactified.
The
toroidally compactified heterotic string exhibits at certain points in
moduli space certain phenomena -- $T$-dualities and extended gauge symmetries
(and gauge groups such as $E_8\times E_8$)  -- that appear to be absent for
Type
I.  One might think that -- if the duality between the heterotic string and
Type I
is correct -- the $T$-dualities must be visible and the extended gauge symmetry
must appear only in a region of moduli space in which the Type I theory is
strongly coupled. A straightforward attempt to implement this idea runs into
trouble at once: the self-dual region of the heterotic string corresponds to a
region of Type I moduli space that seems to intersect weak coupling.

It turns out that the problem is resolved in a somewhat novel and
surprising way.
While the Type I theory appears to be weakly coupled in the relevant region of
its parameter space, we will show that certain of its states become strongly
coupled precisely when the heterotic string exhibits its interesting
world-sheet
behavior. This depends on the unusual properties \ref\polch{J. Dai, R. G.
Leigh,
and J. Polchinski, ``New Connections Between String Theories,'' Mod. Phys.
Lett.
{\bf A4} (1989) 2073.} of the Type I superstring under $R\to 1/R$, which are
related to world-sheet orbifolds considered in\nref\sagnotti{A. Sagnotti,
``Open
Strings And Their Symmetry Groups,'' in Cargese '87, ``Non-perturbative Quantum
Field Theory,'' ed. G. Mack et. al. (Pergamon Press, 1988) p. 521; M. Bianchi
and A. Sagnotti, ``On The Systematics Of Open-String Theories,'' Phys. Lett.
{\bf
247B} (1990) 517.}\nref\horava{P. Horava, ``Strings On World Sheet Orbifolds,''
Nucl. Phys. {\bf B327} (1989) 461, ``Background Duality Of Open String
Models,''
Phys. Lett. {\bf B321} (1989) 251.}
refs.~\refs{\sagnotti,\horava}. Strong coupling of certain states means that
one
cannot exclude the hypothesis that the Type I superstring reproduces the known
heterotic string behavior.  This ``miraculous'' avoidance of a contradiction
adds considerable credence to the idea that these theories really are
equivalent.
This parallels a similar phenomenon in six dimensional heterotic - Type II
duality, where the Type~II theory apparently becomes strongly
coupled at those points in moduli space where the heterotic string has extended
gauge symmetries\nref\asp{P. S. Aspinwall, ``Enhanced Gauge Symmetries and K3
Surfaces,'' preprint CLNS-95/1348, hep-th/9507012 (1995).}\nref\witdyn{E.
Witten,
``Some Comments On String Dynamics,'' preprint IASSNS-HEP-95-63,
hep-th/9507121
(1995).} \refs{\asp,\witdyn}. In fact, we will be
able to see the breakdown of Type~I perturbation theory in a more precise
and detailed way than is presently possible in the Type II case.

Second, pursuing the recent
argument that Dirichlet-branes are intrinsic to the Type I and Type II string
theories and are the carriers of Ramond-Ramond charge \ref\pold{J. Polchinski,
``Dirichlet-Branes and Ramond-Ramond Charges,'' preprint NSF-ITP-95-122, hep-th
9510017.}, we examine the Dirichlet one-brane of the Type I theory
and find that
it has the world-sheet structure of the heterotic string.  This puts the
considerations of \refs{\dab,\hull} on a much firmer footing as
singular solutions of the leading low energy field equations are now replaced
by an exact conformal field theory construction.

\bigskip
\noindent{\it Heterotic String $T$-Duality In Type I Variables}

First we recall the mapping between the low-energy Type I and heterotic string
theories in ten dimensions. Letting $\lambda_h$, $\lambda_I$ be the heterotic
string and Type I coupling constants, and $ g_h$, $g_I$ the two ten-dimensional
metrics, the relations are \witten\
\eqn\uff{\eqalign{ \lambda_I & = {1\over \lambda_h} \cr  g_I & = \lambda_I
g_h\cr}}
(numerical factors are omitted until further notice).

In this paper, we will mainly consider compactification of the ten-dimensional
theory to nine dimensions, on ${\bf R}^9\times {\bf S}^1$. (Some aspects of the
reduction below nine dimensions will also be discussed.) Letting $R_h$,
$R_I$ denote the radius of the circle as measured in the two theories, it
follows
from the second relation in \uff\ that
\eqn\guff{ R_I= \lambda_I^{1/2} R_h.}

The heterotic string world-sheet phenomena whose Type I counterparts we wish to
understand occur for $R_h\leq 1$. Therefore, they will occur for $R_I\leq
\lambda_I^{1/2}$. In trying to see these phenomena at weak Type I coupling --
$\lambda_I\to 0$ -- we will therefore have to take $R_I\to 0$.

\def\onep{{I'}} It will therefore be necessary to understand the small radius
behavior of the Type I superstring. For some string theories, the small $R$
behavior can be understood by an $R\to 1/R$ symmetry. The Type I
superstring does
not have a $T$-duality {\it symmetry}, but one can nevertheless attempt to
understand its small $R$ behavior by means of a $T$-duality {\it
transformation}
to a rather interesting alternative theory that was described in \polch\
(and has
some relations to world-sheet orbifolds discussed  in
\refs{\sagnotti, \horava}). We will call this theory the Type I$'$ theory and
denote the coupling constant and radius as $R_\onep$, $\lambda_\onep$. The
unusual properties of the Type I$'$ theory will be recalled later. For now we
note simply that the Type I and Type I$'$ parameters are related by the
standard
$T$-duality relations \eqn\buff{\eqalign{ R_\onep & = {1\over R_I} \cr
{R_\onep\over \lambda_\onep^2} & ={R_I\over \lambda_I ^2}.\cr}} The second
relation is equivalent to the statement that the nine-dimensional string
coupling
is invariant under $T$-duality.

Combining the above formulas, we see that in the region of $R_h$ fixed and
$\lambda_{I'}\to 0$, $R_{I'}$ will scale as $1/\lambda_\onep$. In
particular, the
$T$-duality of the heterotic string should be visible in a region in which the
Type I$'$ theory is weakly coupled and at large radius -- the region in
which one
thinks one understands it best. This appears to mean that if the heterotic
string
is really equivalent to Type I (and therefore to Type I$'$), the heterotic
string
$T$-duality and extended gauge symmetry should be visible explicitly in the
weakly coupled Type I$'$ theory.

It is easy to make this issue quantitative. Starting with the parameters
$R_\onep, \,\lambda_\onep$ of the Type I$'$ theory, one maps to Type I via
\buff,
and then to the heterotic string by \uff\ and \guff. The composite is
\eqn\unnu{\eqalign{R_h& = {1\over \sqrt{R_\onep \lambda_\onep} } \cr
\lambda_h & = {R_\onep\over\lambda_\onep} \cr  g_h& = g_{I'}{R_{\onep}\over
\lambda_\onep}.\cr}} The last formula is the Weyl transformation of the
nine-dimensional metric deduced from the second equation in \uff. (The
nine-dimensional metric is invariant under $T$-duality between Type I and Type
I$'$ or between the heterotic string and itself.) Then one applies a heterotic
string $T$-duality transformation $R_h\to 1/R_h$, $\lambda_h\to \lambda_h/R_h$,
and finally one inverts \unnu\ to return to Type I$'$ variables. The result is
that the heterotic string $T$-duality corresponds in Type I$'$ to the rather
obscure-looking transformation \eqn\junnu{\eqalign{
R_\onep&\to{R_\onep^{1/4}\over\lambda_\onep^{3/4}}\cr
\lambda_{I'} &\to {1\over \lambda_\onep^{1/4}R_\onep^{5/4}}\cr  g_{I'} & \to
{g_{\onep}\over R_\onep^{1/2}\lambda_\onep^{1/2}}.}}

In particular, the self-dual radius of the heterotic string corresponds to
$R_\onep = 1/\lambda_\onep$, and the region of $\lambda_\onep <<1$ with
$R_\onep
\sim 1/\lambda_\onep$ is mapped to itself, so it appears that we can test in
perturbation theory the existence of this symmetry.\foot{The dual heterotic
theory, which appears at intermediate stages in the transformation of the
Type I$'$ theory into itself, is strongly coupled in this range.  Thus we
are using the fact that its $T$-duality, being a gauge symmetry, is exact
\ref\DHS{M. Dine, P. Huet, and N. Seiberg, ``Large and Small Radius in String
Theory,'' Nucl. Phys. {\bf B322} (1989) 301.}.}

\bigskip\noindent
{\it Transformation Of Masses}

With this in mind, and assuming that the Type I$'$ theory is weakly coupled
when
$\lambda_{I'}<<1$, $R_{I'}>>1$, let us see how particle masses transform under
\junnu.

\def\R{{\bf R}}
\def\S{{\bf S}}
When the Type I theory is formulated on $\R^9\times \S^1$, the particle
momentum
around $\S^1$ is conserved but (as strings in the Type I theory can break) the
string winding number is not conserved. Dually, in the Type I$'$ theory,
winding
number is conserved but momentum is not. Consider therefore a string winding
state in the Type I$'$ theory with a mass of order $R_{I'}$. Heterotic string
$T$-duality, according to \junnu, maps $R_\onep$ to
$R_\onep^{1/4}/\lambda_\onep^{3/4}$. But because of the Weyl transformation in
\junnu, the transformation multiplies masses by an extra factor of $1/R_\onep
^{1/4}\lambda_\onep^{1/4}$, so string winding states with masses of order
$R_\onep$ are exchanged with particles with masses of order $1/\lambda_\onep$.
That is a satisfactory result, as $1/\lambda_\onep$ is the right order of
magnitude for the mass of a Ramond-sector soliton.

The problem arises when one considers the transformation law of other
elementary
string states. For example, a generic Type I$'$ elementary string state has a
mass of order $1$ in string units; though unstable, this state is long-lived
for
sufficiently small $\lambda_\onep$. The heterotic string $T$-duality
transformation should map this state to a long-lived resonance with a mass of
order $1/R_\onep^{1/4}\lambda_\onep^{1/4}$. Such states are not known. Even
worse, consider a (long-lived but not stable) state carrying momentum around
$\S^1$, with a mass of order $1/R_\onep$. Heterotic string $T$-duality would
map
this state to a state with a mass of order $\lambda_\onep^{1/2}/R_\onep^{1/2}$.
Such states, since their mass {\it vanishes} for $\lambda_\onep\to 0$,
would have
to contribute to perturbation theory, where they are not known.\foot{M.
Dine has independently considered the possibility of similar contradictions
arising when the Type I - heterotic and heterotic - Type II dualities are
combined.}

It is suggestive that the masses of the stable particles transform sensibly,
and
that the problem only exists for unstable particles. We will find a solution
compatible with this: the Type I$'$ theory has a strongly coupled region if
$R_\onep $ is of order $1/\lambda_\onep$ or larger, no matter how small
$\lambda_\onep$ may be. The mass formulas for stable particles used above are
exact (because these particles are in ``small'' supermultiplets) and so are
unaffected by strong coupling, but the discussion of unstable particles will be
invalidated by the strong coupling.

\bigskip\noindent
{\it Behavior of the Dilaton}

To see how strong coupling comes about, we must first of all remember that in
string theory, the string coupling constant is determined by the expectation
value of the dilaton field. Stability of the vacuum depends on a cancellation
between dilaton tadpoles. The leading tadpoles come from the disc $D$ and the
projective plane ${\bf RP}^2$, and cancel \ref\gs{M. B. Green and J. H.
Schwarz,
Phys. Lett. {\bf 151B} (1985) 21.} if the gauge group is $SO(32)$.

In the Type I theory, this cancellation occurs homogeneously throughout all
space. In the Type I$'$ theory, however, the situation is rather different,
because of features explained in \polch. Working on ${\bf R}^{9}\times \S^1$,
with $\S^1$ parametrized by a periodic variable $x^{9}$ (of period $ 2\pi
R_\onep$), the Type I$'$ theory is a parameter space orbifold in which a
reversal
of the orientation of the world-sheet is accompanied by $x^{9}\to -x^{9}$. On
the circle, that transformation has two fixed points, at $x^{9}=0$ and
$x^{9}=\pi R_\onep$.

\def\RP{{\bf RP}^2}
Suppose now that $R_\onep$ is large, and consider an ${\bf RP}^2$ mapped to
space-time in such a way that the action is of order one. This means that the
image of $\RP$ in space-time must have a size of order the string scale, not of
order $R_\onep$, and as $\RP$ is unorientable, that is possible only if the
image
of $\RP$ sits near one of the fixed points. There is a symmetry between
these two
points, so the $\RP$ dilaton tadpole receives half its contribution from a
neighborhood of $x^{9}=0$ and half from a neighborhood of $x^{9}=\pi R_\onep$.

Now we come to the tadpole derived from the disc. In going from the Type I
theory
to the Type I$'$ theory, Neumann boundary conditions  are replaced with
Dirichlet
boundary conditions, so the endpoints of the open strings lie at a fixed
position
on $\S^1$. For unbroken $SO(32)$ symmetry (which we assumed on the heterotic
string side to have the $T$-duality that led to the Type I$'$ predictions
that we
are testing), all species of open string must have their boundary at the same
point. The orientifold symmetry $x^{9}\to -x^{9}$ requires that this should be
one of the two fixed points, which we may as well take to be the one at
$x^{9}=0$. The dilaton tadpole from the disc is thus localized near $x^{9}=0$.

The cancellation of dilaton tadpoles therefore occurs in a highly non-local way
in this theory. The disc tadpole near $x^{9}=0$ is only half canceled by an
$\RP$ contribution near $x^{9}=0$; the other half of the cancellation comes
from
$\RP$'s near $x^{9}=\pi R_\onep$. Between the
dilaton source at $ x^{9}=0$, and
the equal and opposite source at $x^{9}=\pi R_\onep$, there is a dilaton
gradient.  The gradient, as it comes from the disk and ${\bf RP}^2$, is of
order $\lambda_\onep$ and so can have an effect of order one when
$R_\onep \cong 1/\lambda_\onep$,
the region where $T$-duality of the heterotic
string is visible.  To first order in $R_\onep \lambda_\onep$ the dilaton is
linear, but (as we will see)
there are nonlinearities whose effect is to make the dilaton
diverge for a finite value of $R_\onep \lambda_\onep$.
The result is to prevent one from comparing
Type I$'$ perturbation theory to heterotic string $T$-duality.

It is also interesting to understand how perturbation theory breaks down
directly in the type I picture, where the radius is becoming small.  Each
additional handle on the world-sheet adds two closed cycles.  The winding
number of $x^9$ is summed for each cycle.  When $R_I$ is small, each sum is
approximated by $R_I^{-1}$ times an integral.  The naive loop expansion
parameter
is therefore $\lambda_I^2 R_I^{-2}$, which is indeed $\lambda_\onep^2$.
However, each additional hole brings an infrared divergence from the
tadpoles of the winding states of the dilaton and graviton.  These do not
cancel winding number by winding number, because $\RP$ contributes only to even
winding numbers while the disk contributes both to even and odd.
The $R_I$-dependence when a hole is added comes only from the cycle which
winds
around the hole, giving a term of order $R_I^2$ in the denominator of the
winding state propagator.  The uncanceled
tadpoles are thus of order $\lambda_I R_I^{-2} = \lambda_\onep
R_\onep$, and so perturbation theory breaks down when this is large.

\bigskip
\noindent{\it Incorporation Of Wilson Lines}

Before making a detailed analysis of the dilaton background, we will
incorporate gauge symmetry breaking by Wilson lines and show how heterotic
string phenomena are mirrored by Type I$'$ phenomena. So we still compactify
the
$SO(32)$ heterotic string on a circle, but now we consider a vacuum in which
the
$x^{9}$ component of the gauge field has an expectation value, with a global
holonomy $W\in {\rm Spin}(32)/{\bf Z}_2$. Regarded as an $SO(32)$ matrix, $W$
can be written as the direct sum of 16 two-dimensional blocks; the $i^{th}$
block takes the form \eqn\ilnsln{\left(\matrix{ \cos \theta_i & \sin \theta_i
\cr  -\sin \theta_i & \cos \theta_i \cr}\right)} with some angle $\theta_i$.
The
incorporation of Wilson lines in the Type I$'$ theory was briefly described in
\ref\other{J. Polchinski, ``Combinatorics Of Boundaries In String Theory,''
Phys.Rev. {\bf D50} (1994) 6041.} and proceeds by shifting the points on the
circle at which strings are permitted to have boundaries. Thus, instead of
requiring as we did above that -- for each of the 32 values of the Chan-Paton
label -- the end of an open string must lie at $x^{9}=0$, one requires that a
charge of type $i$, $i=1,\dots,16$ lies at $x^{9}=\theta_iR_\onep$, while an
image charge of type $\bar i$ lies at $x^{9}=-\theta_iR_\onep$. Here we are
using
a complex basis for the charges; the reality condition for the open string wave
function combines complex conjugation with $x^{9}\leftrightarrow -x^{9}$ and
$i\leftrightarrow \bar i$. As an example that will have some significance, let
$W=W_0$ where $W_0$ has eight $\theta_i$ vanishing and the others equal to
$\pi$. This Wilson line breaks $SO(32)$ to $SO(16)\times SO(16)$. In the Type
I$'$ description, the world-sheet boundary lives at $x^{9}=0$ for 16 values of
the Chan-Paton factor and at $x^{9}=\pi R_\onep$ for the other 16 -- a
configuration constructed by Horava \horava\ as a world-sheet orbifold.

Once a Wilson line is introduced, the heterotic string theory no longer has
$R\to
1/R$ symmetry (which now acts non-trivially on the Wilson line). However, it
still is the case for generic $W$ that as $R$ is decreased, the heterotic
string
theory eventually gets massless particles in a winding sector, giving an
enhanced
gauge symmetry. This phenomenon is invisible in Type I$'$ perturbation
theory, so
to justify our reconciliation of heterotic -- Type I duality with the
predictions
of Type I$'$ perturbation theory, we must show that the Type I$'$ theory
becomes
strongly coupled (somewhere on the ${\bf S}^1$) when the heterotic string would
have enhanced gauge symmetry.

In the bosonic construction of the heterotic string, it is convenient to write
$W=\exp(2\pi i A)$, with $A$ an element of the $SO(32)$ Cartan
 algebra, which is a
copy of ${\bf R}^{16}$. $A$ is only unique up to a shift by an element of the
${\rm Spin}(32)/ {\bf Z}_2$ lattice, and it is convenient to shift $A$ to be as
close to the origin as possible. For instance, in a basis in which the ${\rm
Spin(32)}/{\bf Z}_2$ lattice consists of 16-plets $(m_1,\dots,m_{16})$ that are
all integers or all half-integers and whose sum is even, the group element
$W_0$
described above corresponds to \eqn\guud{A_0=(0,0,\dots,0,1/2,1/2,\dots ,1/2)}
with eight $0$'s and eight $1/2$'s. This vector obeys $A^2=2$ (and cannot be
shifted by a lattice vector to be closer to the origin). A small exercise shows
that any vector in $\R^{16}$ that is not equivalent to $A$ up to a lattice
shift
is (up to a lattice shift) closer to the origin than $A$.

In the absence of a Wilson line, the heterotic string gets at the self-dual
radius enhanced
$SU(2)$ gauge symmetry, due to a massless state with unit momentum and winding
around $x^{9}$. In the presence of the Wilson line, the enhanced gauge symmetry
occurs at a radius that is smaller by a factor of $\sqrt{1-A^2/2}$; this is
essentially because a shift by $A$ increases the left-moving ground state
energy
by $A^2/2$. Thus, for any Wilson line other than $W=W_0$, the heterotic string
gets enhanced gauge symmetry if $R$ is small enough.

Since this enhanced gauge symmetry will not be visible in Type I$'$
perturbation
theory, we must hope that the Type I$'$ theory becomes strongly coupled for
sufficiently big radius unless $W=W_0$. This is so. For
$W=W_0$, the open string boundaries are at $x^{9}=0$ for half of the values of
the Chan-Paton factor, and at $x^{9}=\pi R_\onep$ for the other half. Thus half
of the dilaton tadpole from the disc is localized at one orientifold fixed
point
and half at the other. This is the same configuration as for $\RP$'s, so the
cancellation of tadpoles occurs locally, and there is no linear dilaton field
induced. One can therefore take $R_\onep\to\infty$ with no breakdown of
Type I$'$
perturbation theory. For any other value of $W$, dilaton tadpoles
cancel between
different points on the circle, and for sufficiently big $R_\onep$ --
corresponding to sufficiently small heterotic string radius -- one gets strong
coupling somewhere on the circle. Thus, the occurrence of strong coupling
for
Type I$'$ mirrors in a striking fashion the occurrence of extended gauge
symmetry
of the heterotic string.

%
%

\bigskip
\noindent{\it Type I$'$ Dilaton Background}

This discussion can be made much more quantitative by using the
equations of supergravity to solve for the low energy fields in the
situation studied above.  In this section we are careful to retain all
numerical factors.

First,  let us write the low energy field theory of the ten
dimensional $SO(32)$ heterotic string, keeping only the graviton, dilaton, and
gauge fields.  The action is
\eqn\hetact{ S_{h} = \int d^{10}x \, \sqrt{-g} e^{-2\phi} \left(
{1\over 2} R+ 2 \del_M \phi \del^M\phi - {1\over 8} {\rm Tr}_V G_{MN} G^{MN}
\right)}
The overall normalization of the action and the relative normalization of the
gauge kinetic term have been fixed by choice of the additive normalization of
the dilaton and the multiplicative normalization of the metric.  In particular,
this normalization of the metric corresponds to setting $\alpha' = 2$
\ref\het2{D. J. Gross, J. A. Harvey, E. Martinec, and R. Rohm,
``Heterotic String Theory (II),'' Nucl. Phys. {\bf B267} (1995) 75. }.

Let us similarly write the action for the Type I$'$ string with one dimension
compactified on $0 \leq x^9 \leq 2\pi$, the endpoints being orientifold points.
The indices $M,N$ run $0 \ldots 9$ and the indices
$\mu,\nu$ run $0 \ldots 8$.
We include also 8-branes perpendicular to the 9-direction, with $n_1$
8-branes at
$x^9_1$, $n_2$ at $x^9_2$, and so on (and their images at $-x_i^9$).
As explained earlier, the positions of the 8-branes are related to the
Wilson lines in  a Type I description.
Initially
let $0 < x^9_i < 2\pi$, so the gauge group is $U(n_1) \times U(n_2) \times
\ldots$.  We keep the same fields as above plus the 9-form potential $A$
\pold, which will also have a nontrivial background.
The action is
\eqn\onepact{ \eqalign{ S_{I'} = \int d^{10}x& \, \sqrt{-g} e^{-2\phi} \left(
{1\over 2} R+ 2 \del_M \phi \del^M \phi \right)  - {1\over 2}
\int F^*F \cr
& - \mu_8 \sum_i \int_{x^9 = x^9_i} \left( d^9 x\, \sqrt{-\tilde g} e^{-\phi}
2^{-1/2}
\left\{ n_i + (\pi\alpha')^2 {\rm Tr}_f G_{\mu\nu} G^{\mu\nu} \right\} + n_i A
\right) }}
where $\tilde g_{\mu\nu}$ is the 9-dimensional metric.
Again, the additive normalization of $\phi$ is fixed by the gravitational and
dilaton kinetic terms, while the normalization of $F = dA$ is defined by its
kinetic term.  The coupling of the 9-form potential to the 8-brane is as in
ref.~\pold, $\mu_8 = (2\pi)^{-9/2} \alpha'^{-5/2}$.  The normalization of the
membrane tension is fixed by supersymmetry, as we will see below.  The
normalization of the gauge kinetic term then follows from the Born-Infeld form
of the open string action~\ref\bornin{E. S. Fradkin and A. A. Tseytlin, Phys.
Lett. {\bf B163} (1985) 123.}.  There is no particularly simple choice for the
Type I$'$ $\alpha'$ so we leave it arbitrary.

The equation of motion
of the nine-form potential implies that the ten-form
is  $F = \nu_0 dx^0 \ldots dx^9$, with $\nu_0$ piece-wise constant away from
the
8-branes. At $x^9_i$ the equations of motion imply a jump
$\Delta \nu_0 = n_i \mu_8$.  Since there are 16 8-branes between the two fixed
points, we have
$\nu_0(2\pi) = \nu_0(0) + 16 \mu_8$.  Invariance of the boundary conditions
under spacetime parity $x^9 \to 2\pi -  x^9$, which takes $\nu_0 \to - \nu_0$,
then implies that
$\nu_0(2\pi) = - \nu_0(0) = 8\mu_8$. Boundary conditions on the metric and
dilaton will be seen to follow from supersymmetry.  The case that some of the
$x^9_i$ are at the orientifold points $0$ or $2\pi$ can now be approached as a
limit.  The one qualitative change is that some vectors, which wind between an
8-brane and its image, become massless so that $U(n)$ is promoted to $SO(2n)$.
The traces in the $U(n)$ fundamental and
$SO(2n)$ vector are related by ${\rm Tr}_f = {1\over 2}{\rm Tr}_V$.

To solve for the background fields, look first in the region between the
8-branes, where as explained in ref.~\pold\ the theory reduces to the
massive IIa
supergravity found by Romans~\ref\romans{L. J. Romans, ``Massive $N=2a$
Supergravity in Ten Dimensions,'' Phys. Lett. {\bf B169} (1986) 374.}.  The
action, metric and dilaton of that paper are related to those here by
$S^{[15]} = - {1\over 2} S$, $g^{[15]}_{\mu\nu} = e^{-\phi/2} g_{\mu\nu}$,
$\phi^{[15]} = -{1\over 2} \phi$, and the parameter $m$ of that paper is
$\nu_0\sqrt 2$.  The conditions for a supersymmetric background, in terms of
the variables of ref.~\romans, are
\eqn\susyvar{\eqalign{
32 D_M \epsilon &= m e^{-5\phi/2} \Gamma_M \epsilon \cr
8\partial_M \phi \Gamma^M\epsilon &= 5 m e^{-5\phi/2} \epsilon .}}
The 8-branes and orientifolds preserve half the supersymmetries, namely
those with
$\Gamma_9 \epsilon = \pm g_{99}^{1/2} \epsilon$; the sign, which is
correlated with the charge of the 8-brane, will be determined below.
Integrating eq.~\susyvar\ and transforming from the variables of ref.~\romans\
to those used in the rest of this paper gives, in conformal gauge $g_{MN} =
\Omega^2(x^9)
\eta_{MN}$,
\eqn\soln{\eqalign{
&e^{\phi(x^9)} = z(x^9)^{-5/6}, \qquad \Omega(x^9) = C z(x^9)^{-1/6} \cr
& z(x^9) = 3 C (B \mu_8 \pm \nu_0 x^9) /\sqrt 2  }}
with $C$ and $B$ piecewise constant between the 8-branes.

This background
satisfies the equations of motion between the 8-branes.  At the 8-branes, the
equations of motion require $\phi$
and $\Omega$ to be continuous and $\nu_0$ to have the discontinuity $\mu_8$.
This implies that $C$ is constant, that $z(x^9)$ is continuous, and therefore
that
$B$ has a discontinuity $\mp n_i x^9_i$; the full
$x^9$-dependence of
$B$ is then known given $B(0)$.  One can now work backwards and determine the
normalization of the membrane tension.  From the solution \soln, the
discontinuity in $\del_9 \phi$ is $\mp 5C n_i \mu_8/2\sqrt{2}z$.
Relating this to the dilaton equation of motion gives the value
$\mu_8/\sqrt{2}$ used in the action \onepact.  Also, positivity of the
membrane tension determines that we must take the lower sign in $z(x^9)$.
Finally the solution~\soln\ determines the $\phi$ and $\Omega$ gradients in
terms of $\nu_0$, thus giving the boundary conditions on these at the
orientifold points as promised earlier.

This Type I$'$ theory is supposed to be dual to the heterotic theory quantized
on a circle of radius $R$.  To find the relation between the heterotic $R$ and
$\phi$ and the parameters $B(0)$ and $C$ of the Type~I$'$ theory we compare the
effective 9-dimensional actions obtained by reducing the respective
10-dimensional actions~\hetact\ and~\onepact.  For the
Type~I$'$ metric we assume a slow function $\gamma_{\mu\nu}$ of the noncompact
dimensions times the $\Omega^2(x^9)$ determined above.  We also need to
determine the relation between the metrics, $\gamma_{\mu\nu} = D^2
g_{\mu\nu}^{h}$.  Comparing the gravitational actions gives
\eqn\grav{
2\pi R e^{-2\phi} = D^7 C^{25/3} \int_0^{2\pi} dx^9\,
w(x^9), }
where $w(x^9) = 3^{1/3} 2^{-1/6} [\mu_8 B(x^9) - x^9 \nu_0(x^9)]^{1/3}$.
Comparing the gauge actions gives
\eqn\gauge{
2\pi R e^{-2\phi} = (2\pi\alpha')^2 2^{-1/2} \mu_8 D^5 C^5. }
A final relation is obtained by comparing the mass of a BPS state, a
heterotic Kaluza-Klein state which is dual to a Type I$'$ winding state.
We have
$m_{h} = 1/R = D m_{I'}$.  Integrating the world-sheet action gives
\eqn\mass{
{1 \over R} = {D C^{5/3} \over \pi \alpha'}
\int_0^{2\pi} dx^9\, w(x^9)^{-1}\ . }
In deriving this note that the
string winds from 0 to $2\pi$ and back again, and that the area element for
the winding state is $\Omega^2 dx^0 dx^9$.
The relation between the heterotic Wilson line and the
positions of the 8-branes can similarly be found by considering the mass of
an off-diagonal vector boson, $(A_i \pm A_j)/R$ in the heterotic string.
In the Type I$'$ theory this is a string attached to the branes at $x^9_i$ and
$x^9_j$ (or the image at $-x^9_j$).  Thus,
\eqn\wilmap{
{\lambda_i \over R} = {D C^{5/3} \over 2\pi \alpha'}
\int_0^{x^9_i} dx^9\, w(x^9)^{-1}\ . }

We can combine eqs.~\grav, \gauge, and~\mass\ to get
\eqn\rad{
R = 2^{-3/4} \mu_8^{-1/2}
\left( \int_0^{2\pi} dx^9\, w(x^9) \right)^{1/2}
\left( \int_0^{2\pi} dx^9\, w(x^9)^{-1} \right)^{-1}  }
We will focus here on the simple case that $A$ has $n$
0's and $16-n$ ${1\over 2}$'s so there are $2n$ branes at
$x^9=0$ and $32-2n$ at $x^9 = 2\pi$ (counting the images); let $n \leq 8$ for
convenience.
In this case $\nu_0(x^9) = (n-8)\mu_8$ and $B(x^9) = B$ are constants, and
\eqn\dil{ \eqalign{
&e^{\phi(x^9)} =
\left\{ 2^{-1/2} 3 C \mu_8 [B + (8-n) x^9 ] \right\}^{-5/6} \cr
&R = {1\over 2} 3^{1/2}
\left( \int_0^{2\pi} dx^9\, [B + (8-n) x^9 ]^{1/3} \right)^{1/2}
\left( \int_0^{2\pi} dx^9\, [B + (8-n) x^9 ]^{-1/3} \right)^{-1}
}  }
When $B$ is large, the coupling $e^{\phi(x^9)}$ is small everywhere and the
Type I$'$ theory is weakly coupled; in this range the heterotic radius $R$ is
large.  However, when $B \to 0$ the coupling diverges at the endpoint
$x^9=0$ and
perturbation theory breaks down;\foot{It diverges as the inverse of the proper
distance, in the type I$'$ string metric, from the endpoint.}
 past this point the expressions make no
sense.  Evaluating the integrals for $B=0$ gives the point of breakdown as
\eqn\breakdown{
R = {1\over 2} |n-8|^{1/2}. }
This has been extended to $n > 8$ by symmetry; the coupling diverges at the
orientifold point with the fewer D-branes.  For each $n$ this is the precise
radius at which an enhanced gauge symmetry appears on the heterotic side.  Thus
the conjectured heterotic -- Type I duality ``miraculously'' survives
confrontation with heterotic string $T$-duality.  It would be interesting to
extend this analysis to other enhanced symmetry points, such as the $E_8
\times E_8$ point.

\bigskip\noindent {\it Compactification Below Nine Dimensions}

Let us now consider the heterotic string compactified below nine dimensions in
the light of conjectured duality with Type I. Regardless of the dimension, the
relation $R_I=\lambda_I^{1/2}R_h$ shows that if $R_h$ is of order one (so that
the heterotic string has interesting world-sheet dynamics) and $\lambda_I$ is
small (so that Type I perturbation theory is useful), then $R_I$ will be small,
so that the Type I description is difficult to interpret. We therefore consider
instead the dual Type I$'$ description,\foot{One can also consider a mixture of
Type I and Type I$'$, dualizing in some directions only; this seems unlikely to
raise essentially new issues.} and must show in each dimension that if one
starts
with a heterotic string at large radius and moves in towards small radius, Type
I$'$ perturbation theory breaks down by the time the heterotic string has
interesting dynamics.

We begin with the heterotic string on $\R^8\times \S^1\times \S^1$. For
simplicity we ignore the $B$ field and consider the two $\S^1$'s to be
orthogonal
with respective radii $R_{h,1}$ and $R_{h,2}$.

Something new is needed, for the following reason. As long as $R_{h,1}$ and
$R_{h,2}$ are of the same order of magnitude, a clash between heterotic string
$T$-duality and Type I$'$ perturbation theory cannot be avoided by generating a
linear dilaton. Indeed, with two compact dimensions (of roughly the same size)
the solution of the Laplace equation for a dilaton with a source will grow at
most logarithmically, not linearly, and this will not help much.

Happily, in eight dimensions the transformation law from the heterotic string
to
Type I$'$ is significantly different from what we have worked with in nine
dimensions. The change is in the $T$-duality transformation from Type I to Type
I$'$. One still has the usual transformation law of the radii,
$R_{I',i}=1/R_{I,i}$. But as the eight-dimensional string coupling constant
is to
be invariant under $T$-duality, the ten-dimensional coupling transforms as
\eqn\huddo{{R_{\onep,1}R_{\onep,2}\over \lambda_\onep^2} ={R_{I,1}R_{I,2}\over
\lambda_I^2}.} Together with the map $R_{I,i}=R_{h,i}\lambda_I^{1/2}$ between
heterotic and Type I variables, this implies the perhaps surprising formula
\eqn\surprising{\lambda_\onep={1\over R_{h,1}R_{h,2}}.} If therefore the
$R_{h,i}$ are of the same order of magnitude, and are of order one so that the
heterotic string has interesting world-sheet dynamics, then $\lambda_\onep$
will
be of order one, so that Type I$'$ perturbation theory is not useful. One could
try to avoid this by taking, say, $R_{h,1}$ of order one, so that the heterotic
string has interesting world-sheet dynamics, and $R_{h,2}$ large so that
$\lambda_\onep$ is small. In this case, the torus of the Type I$'$ theory is
highly anisotropic and quasi one-dimensional at big distances; one will again
meet a linear dilaton, driving one to strong coupling at some point on the
torus
even though the bare coupling given in \surprising\ is small.

Below eight dimensions, the details are again slightly different. Upon
compactification of the heterotic string on an $n$-torus of $n>2$ with radii
$R_{h,i}$, the Type I$'$ radii turn out to be given by
\eqn\hujju{R_{I',i}^{n-2}=\lambda_{I'}\prod_{j=1}^nR_{h,j}\cdot {1\over
R_{h,i}^{n-2}}.} For simplicity, suppose that $k$ of the $R_{h,j}$ are of order
one and $n-k$ are of the same order $R>>1$. Also suppose $\lambda_\onep <<1$ or
Type I$'$ perturbation theory is in any case not a good description.
For $k=0$, there is no interesting
dynamics of the heterotic string. For $k=1$, one of the $R_{I'}$ is much
than the others by a factor of $R$ and -- taking $\lambda_{I'}R\sim 1$ or
bigger
so that none of the $R_{I',i}$ is smaller than the string scale -- the Type
I$'$
theory is strongly coupled because of a linear dilaton. For $k>1$, some of the
$R_{I'}$ are $<<1$ and again the Type I$'$ description becomes untransparent.

\bigskip\noindent
{\it Wilson Lines Below Nine Dimensions}

It is also of interest to consider gauge symmetry breaking in compactification
below nine dimensions. Consider, for example, the heterotic string on
$\R^8\times
\S^1\times \S^1$ with periodic coordinates $x^8,\,x^{9}$ and Wilson lines
$W_1,$
$W_2$. We will consider only the dependence on the metric of $\S^1\times \S^1$,
as the $B$-field corresponds on the Type I$'$ side to a little-understood
Ramond-Ramond modulus.

For what choices of $W_1,$ $W_2$ does the heterotic string {\it never} get
extended gauge symmetry, for {\it any} choice of the flat metric on $\S^1\times
\S^1$? This question is easily answered. It is necessary that $W_1$ should be
conjugate to the matrix $W_0$ introduced above (which breaks $SO(32)$ to
$SO(16)\times SO(16)$), or a state with momentum and winding in the $x^8$
direction would become massless when the first circle is small. $W_2$ must be
in
the same conjugacy class, or a state with momentum and winding in the $x^{9}$
direction would become massless when the second circle is small. And the
product
$W_1W_2$ must be in the same conjugacy class, or a state with equal momentum
and
winding in the two directions would become massless when the metric is small in
the diagonal direction. These conditions imply in particular that
$W_1^2=W_2^2=(W_1W_2)^2=1$ and therefore that $W_1$ and $W_2$ commute and so
can
be simultaneously diagonalized. The condition on the conjugacy classes now
implies that the states of $W_1=1$ (and likewise the states of $W_1=-1$) are
equally divided between $W_2=1$ and $W_2=-1$, and hence that $W_1$ and $W_2$
together break $SO(32)$ down to $SO(8)^4$.

Let us compare this to what happens on the Type I$'$ side. On $\S^1\times \S^1$
there are $2\times 2=4$ orientifold fixed points, each bearing one fourth of
the
$\RP$ dilaton tadpole. To avoid getting a large growth of the dilaton {\it
regardless} of the metric on $\S^1\times\S^1$, the cancellation of the dilaton
tadpoles must occur locally; eight values of the Chan-Paton index must be
supported at each of the four fixed points. This therefore breaks $SO(32)$ to
$SO(8)^4$, showing that the Type I$'$ theory never gets to strong coupling
precisely if the Wilson lines are such that the heterotic string never has
extended gauge symmetry.

One can straightforwardly extend this below eight dimensions. Upon toroidal
compactification to $10-n$ dimensions, the heterotic string never gets enhanced
gauge symmetry, and the Type I$'$ theory never gets strong coupling because of
a
linear dilaton, precisely if the Wilson lines are such as to break $SO(32)$ to
$SO(2^{5-n})^{2^n}$. Of course, for $n>5$ this is impossible.

\bigskip{\it The Type I Dirichlet One-Brane}

In the remainder of this paper, we will approach the question of
heterotic - Type I duality in a quite different way.  We consider
the D-string of the Type I theory and show that it has the world-sheet
structure of the heterotic string, thus refining earlier discussions
\refs{\dab,\hull} based on singular approximate classical solutions.
The D-string has a tension in Type
I theory of order $1/\lambda_I$, an expression that is exact
because of BPS saturation.  The D-string therefore becomes very light
when the Type I theory is strongly coupled, and it is very
plausible -- given
the world-sheet structure -- that it then behaves like an elementary
heterotic string.  As a check, recall that string tension has dimensions
of mass squared and so is multiplied by $\lambda_I$ under the Weyl
transformation $g_I=\lambda_Ig_h$ between Type I and
heterotic string metrics;
hence the D-string has a tension of order one in
the heterotic string metric,
like the elementary heterotic string.

Much of the world-sheet structure of the heterotic string can
be anticipated just from considerations of supersymmetry.
Consider  a D-string that is located at, say, $ x^2=\dots = x^9=0$.
This configuration is invariant under a subgroup
$SO(1,1)\times SO(8)$ of the Lorentz group $SO(1,9)$,
where $SO(1,1)$ acts on $x^0,x^1$ and $SO(8) $ on $x^2,\dots,x^9$.
The ten-dimensional supersymmetries transform as ${\bf 16}$ of $SO(1,9)$,
which decomposes as ${\bf 8}'_+\oplus  {\bf 8}''_-$ of
$SO(1,1)\times SO(8)$, with ${\bf 8}'$, ${\bf 8}''$ the
two spinor representations of $SO(8)$, and $\pm $ the $SO(1,1)$
charge.  In the field of the D-string, half the supersymmetry,
say ${\bf 8}'_+$, survives.  The broken ${\bf 8}''_-$ instead
generates chiral fermions zero modes along the world-sheet,
as discussed in \ref\zeroworld{J. Hughes and J. Polchinski,
``Partially Broken Global Supersymmetry And The Superstring,''
Nucl. Phys. {\bf B278} (1986) 147.}.
Together with eight bosonic zero modes associated with  the broken
translational symmetries, these eight chiral fermion zero modes
are the heterotic string world-sheet degrees of freedom
that carry space-time quantum numbers.  To reproduce the heterotic
string, one also must find current algebra degrees of freedom
of appropriate chirality.

To actually compute the excitation spectrum of the D-string,
one simply quantizes the open string allowing Dirichlet
ends with $x^2=\dots = x^9=0 $ as well as the standard free
string Neumann boundary conditions.  There are thus two
new open string sectors to consider, the DD sector with
Dirichlet boundary conditions at each end, and the DN sector
with Dirichlet boundary conditions at one end and Neumann at the other.
Each sector, of course, can be further subdivided into Neveu-Schwarz
and Ramond sectors.

We consider the DD sector first, in covariant RNS formalism.
World-sheet bosons and fermions $X^2,\dots,X^9$ and $\psi^2,\dots,
\psi^9$  get an extra minus
sign in reflection from the boundary, as a result of replacing
Neumann by Dirichlet, but this happens twice, so (in either the
Ramond or Neveu-Schwarz sector) the world-sheet bosons and fermions
have the standard integrally or half-integrally moded expansions,
giving the usual ground state energies and hence apparently leading
to  a  massless spectrum consisting of the usual
vector supermultiplet.
But the zero modes of $ X^2,\dots,X^9$ are absent, so that the
massless DD (or DN) modes are functions of $x^0,x^1$ only;
that is, they propagate only on the D-string world-sheet.
Thus, the ten-dimensional massless vector $A_I$ becomes in this
context a world-sheet vector $A_i(x^0,x^1)$, $i=1,2$ and scalars
$\phi_j(x^0,x^1)$, $j=2,\dots, 9$.

Also, differences from usual open string quantization arise when
we consider the fact that the Type I string is {\it unoriented},
so that the DD spectrum must be projected onto a subsector invariant
under the operator $\Omega$ that  exchanges
the two ends.  The vector $A$ has a conventional
vertex operator $V_A=\sum_{i=0,1} A_i\,{\partial
X^i/ \partial\tau}$, with $\partial_\tau$
the derivative tangent to the boundary.  $V_A$ is odd under reversal
of orientation of the boundary, so the vector is projected out
-- as usually occurs for open strings without Chan-Paton factors.
But as briefly explained in \other, the scalars  have vertex
operator $V_\phi=\sum_{j=2}^9\phi_j\,{\partial X^j / \partial\sigma}$, with
$\partial_\sigma$ the {\it normal} derivative; this is even
under reversal of the boundary orientation, so that the scalars
survive the projection.  They indeed represent the oscillations in
position of the D-string.  Now consider the action of  $\Omega$
on massless fermions.  In the Ramond sector, the RNS fermions
$\psi^I$ have zero modes $\Gamma^I$ that upon quantization
obey $\{\Gamma^I,\Gamma^J\}=2\eta^{IJ}$.  The GSO projection restricts
the massless fermions to states invariant under an operator $(-1)^F$
that anticommutes with all $\psi^I$; in the space of zero modes
this operator can be represented by $\bar\Gamma=\Gamma^0\Gamma^1\dots
\Gamma^9$.  For standard NN open strings,
the operator $\Omega$ can be taken to commute with
the $\Gamma^I$ (otherwise replace $\Omega$ by $\Omega\cdot (-1)^F$,
since one is in any case making the GSO projection), but  turns out
to act as $-1$ on the ground state, which therefore altogether disappears
from the spectrum, a standard result
for unoriented open strings without Chan-Paton factors.
Now consider the DD case.  Because the
fermions $\psi^2,\dots,\psi^9$ reflect from the boundary with
an extra minus sign compared to $\psi^0,\psi^1$, they pick up
an extra minus sign under exchange of left and right movers.  Instead
of being represented on massless fermions by $-1$, $\Omega$ therefore
acts by $\Omega = - \Gamma^2\Gamma^2\dots \Gamma^9$ (or
by $-\Gamma^0\Gamma^1$ if it is multiplied by a factor of
$(-1)^F$).  Physical massless fermions are thus spinors $\chi$
that obey
\eqn\jury{\chi = \bar\Gamma\chi=-\Gamma^2\Gamma^3\dots\Gamma^9\,\chi.}
The first condition says that $\chi$ transforms in the ${\bf 16}$ of
$SO(1,9)$, and the second that in the decomposition
${\bf 16}={\bf 8}'_+\oplus {\bf 8}''_-$, $\chi$ transforms as
${\bf 8}''_-$.  The $-$ $SO(1,1)$ charge
 means that $\Gamma^0\Gamma^1\chi
=-\chi$, so that $\chi$ is right-moving on the world-sheet.

Thus, the DD sector gives precisely the
 massless world-sheet modes of the
heterotic string, except the current algebra
 modes, which  must come from the DN sector.
In this sector, the world-sheet bosons $X^2,\dots ,X^9$ get
an extra minus sign (compared to ordinary open strings)
in reflection at the Dirichlet end and so have
a {\it half-integral} mode expansion.  The fermions $\psi^2,\dots,
\psi^9$ similarly have an extra minus sign, so have
 an {\it integer} mode expansion in the Neveu-Schwarz sector
and a {\it half-integer} expansion in the Ramond sector.  These
facts mean that in the  Neveu-Schwarz sector, the ground state
oscillator energy is strictly positive, and there are no massless
states.  The massless spectrum consists therefore only  of fermions.
In the Ramond sector, the ground state energy is zero and
the only  fermion zero modes are $\Gamma^0$ and $\Gamma^1$,
whose quantization gives two states, of which only the
left-moving mode survives the GSO projection
$\Gamma^0\Gamma^1\lambda=\lambda$.
Because $\lambda$ carries also the Chan-Paton factors of the Neumann
boundary, one gets the expected current algebra modes of the
heterotic string -- 32 left-moving world-sheet fermions transforming
in the ${\bf 32}$ of $SO(32)$.  Space-time supersymmetry of course requires
that $\lambda$ and $\chi$ have opposite chirality.

We have considered the Type I GSO projection, but there are also GSO
projections on the heterotic side, on the current algebra fermions and on the
$\psi^I$.  We should find the corresponding projections on the
D-string spectrum.  The GSO projection on the supersymmetric fermions is
automatic because the DD fermions are spacetime spinors, Green-Schwarz
fermions.  For the current algebra fermions there is a natural origin for the
GSO projection. We have noted that the world-sheet $U(1)$ gauge field is
removed by the orientation projection.  A ${\bf Z}_2$ subgroup,
holonomies $\pm 1$, commutes with the orientation reversal, and a consistent
string theory is obtained for any choice of this holonomy around closed loops
on the D-string world-sheet.  The nontrivial holonomy gives a $-1$ for each D
endpoint, and so acts trivially on the DD strings but acts as $-1$ on the DN
strings.  Treating this ${\bf Z}_2$ as a discrete gauge group generates
the GSO projection on the current algebra fermions.
Evidently the rules of D-branes require us to sum over all consistent
theories in this way, a result which will be relevant for counting D-brane
states in other contexts.

\bigskip\noindent
{\it D-Branes And Solitons}

To further compare the heterotic and Type I theories, one might
ask whether other Type I D-branes can be identified in
the heterotic string.
Actually, the only supersymmetric Type I D-brane other
than the D-string is the five-brane.  Its tension is of
order $1/\lambda_I$, and allowing for the effects of the Weyl
transformation, it corresponds to a heterotic string five-brane
with a tension of order $1/\lambda_h^2$.  This is the standard
behavior of the solitonic five-brane of the heterotic string
\nref\strom{A. Strominger, ``Heterotic Solitons,''
Nucl. Phys. {\bf B343} (1990) 167.}
\nref\cal{C. G. Callan, Jr., J. A. Harvey,
and A. Strominger, ``World-Sheet Approach To Heterotic
Instantons And Solitons,'' Nucl. Phys. {\bf B359} (1991) 611.}\nref\dlk{M. J.
Duff, R. R. Khuri, and J. X. Lu, ``String Solitons,'' Phys. Rept. {\bf 259}
(1995) 213-326,1995, hep-th/9412184. }\refs{\strom - \dlk}, so it is natural to
identify the two.

Incidentally, this is apparently not the only example of
a D-brane that transforms into an ordinary soliton under
string-string duality.  Consider the relation between the Type IIA
theory on ${\bf R}^6\times {\rm K3}$ and the
heterotic string on ${\bf R}^6\times {\bf T}^4$.
 The Type IIA six-brane, wrapped around
K3, becomes a two-brane  in ${\bf R}^6$, with tension of order
$1/\lambda_{IIA}$.  Under string-string duality, allowing
for the Weyl transformation, it transforms into a two-brane
with a tension of order $1/\lambda_h^2$, plausibly visible
as an ordinary soliton.  There is a very
natural candidate for what this two-brane is.  Indeed, Johnson,
Kaloper, and Khuri
\ref\johnson{C. V. Johnson, N. Kaloper, and R. R. Khuri,
``Is String Theory A Theory Of Strings?'' hep-th/9509070.}
described a heterotic string solitonic two-brane on
${\bf R}^6$ (related to magnetic black holes in four-dimensions)
that transforms under heterotic - Type II duality into a Type IIA
two-brane that classically appears to be singular; this is very
plausibly the Dirichlet two-brane.

\bigskip

We would like to thank M. Dine and A. Strominger for discussions.
This work was supported in part by NSF grants PHY91-16964, PHY92-45317,
and PHY94-07194.

\listrefs
\end